\documentclass[prb,twocolumn,showpacs,superscriptaddress,groupedaddress]{revtex4-1}
\usepackage{amsfonts}
\usepackage{amsmath}
\usepackage{amssymb}
\usepackage{graphicx}
\usepackage{graphicx,amssymb,amsbsy,color}
\usepackage{bm} 
\usepackage[dvipdfm,colorlinks=true,linkcolor=blue,citecolor=blue,urlcolor=blue]{hyperref}

\begin{document}

\title{Consequences of extended $s_\pm$-wave pairing in iron-based superconductors}
\author{Hsuan-Hao Fan$^{1}$, C. S. Liu$^{2}$, and W. C. Wu$^{1}$}
\affiliation{$^1$Department of Physics, National Taiwan Normal
University, Taipei 11677, Taiwan\\
$^2$Department of Physics, Yanshan University, Qinhuangdao 066004, China}

\date{\today}

\begin{abstract}
Motivated by a recent experiment of Song \emph{et al.} [Science {\bf
332}, 1410 (2011)], we theoretically study the spin dynamics, charge
dynamics, and point-contact Andreev-reflection spectroscopy (PCARS)
of two-band iron-based superconductors of a possible
extended $s_\pm$-wave pairing symmetry. We consider the case of a
dominant $s_\pm$ gap blended by a secondary extended $s$ component
in which gap nodes can develop in the Fermi pockets near
zone corner and/or boundary. Due to the strong nesting effect associated
with nodal regions, dynamical spin and charge susceptibilities can exhibit
strong peaks at momenta near $(\pm\pi/2,0)$, $(\pm\pi,\pm\pi/2)$,
as well as $(\pm\pi,0)$ in the unfolded Brillouin zone.
For PCARS, considering an anisotropic band effect induced by
an applied voltage, [100] differential conductance can exhibit a $V$-shape
behavior manifesting a gap node occurring in such direction. It is highly
suggested that the above features can be experimentally investigated
to help sorting out the pairing symmetry of iron-based superconductors.
\end{abstract}
 \maketitle

\section{Introduction}\label{sec1}

Pairing symmetry of the Fe-based pnictide and chalcogenide
superconductors is currently a hot topic in the study of
superconductivity. While more and more experimental results have
suggested that the order parameter in these materials is likely to
be fully gapped $s_\pm$-wave,
\cite{Umezawa2012,Chen2010,Hanaguri2010,Hashimoto2009} whether there
is a node in the gap remains controversially.\cite{Hicks2009,Dong2010}
For instance, in a recent scanning tunneling microscopy (STM) measurement of
Song \emph{et al.},\cite{Song2011} a nodal and two-fold symmetry gap is revealed
in an iron selenide (FeSe) superconductor. It is no doubt that a central issue
towards understanding the iron-based superconductors (FeSCs) is to unambiguously
identify the pairing symmetry of these materials.

Among many different probes, inelastic neutron scattering (INS) measures the
two-particle excitations and can give direct information on the momentum
and energy dependence of the quasiparticle excitation and the pairing gap. A strong
coherence peak can emerge in the dynamic spin susceptibility if the corresponding
two-particle excitation is highly degenerate ({\rm i.e.}, in good nesting condition).
In addition, strong peaks in INS can occur due to the resonant nature.
A conclusive INS measurement of FeSCs remains unsettled however to
which wave vectors of the strong peaks are observed to be remarkably material
dependent. Some INS measurements have reported that spin resonances
occur at the wave vector near $(\pi,\pi)$ [or $(\pi,0)$] in the folded (or unfolded)
Brillouin zone (BZ) in 1111,\cite{Shamoto2010} 122,\cite{Christianson2008,Ishikado2011}
111,\cite{Taylor2011} and 11 families.\cite{Qiu2009,Mook2010} Other
INS measurements on the new 122* family $\text{A}_x\text{Fe}_2\text{Se}_2$
(A$=$K, Rb, and Cs)\cite{Liu2011} of electron Fermi
surfaces only\cite{Qian2011,Zhang2011} have revealed that a
resonance peak occurs at the wave vector $(\pi,\pi/2)$ in the
unfolded BZ.\cite{Friemel2012} On the theoretical side, on the other
hand, prediction of the positions of the resonance peak in momentum
space remains controversial. It has been predicted that for $s_\pm$ pairing
state, a strong coherence peak can exist in the dynamic spin susceptibility
$\chi(${\bf Q}, $\omega)$ at the nesting wave vector {\bf Q}=$(0,\pi)$ or
$(\pi,0)$ in the unfolded BZ.\cite{Mazin2008,Korshunov2008,Maier2008} However, based on
a pairing potential associated with the predicted $\chi(${\bf Q}, $\omega)$, the
SC state are found to be inconsistent with the previous prediction.\cite{Das2011,Maier2011}

Charge dynamics which is accessible by high-energy electron scattering or X-ray scattering
is another ideal candidate for studying the pairing symmetry of iron-based
superconductors. While charge and spin susceptibilities are coupled to different
coherence factors, as far as two-particle excitation is concerned they are qualitatively similar.
Thus for a comparison point of view, it is also useful to theoretically study the
charge dynamics within the same framework.

Another high-resolution phase-sensitive probe to detect the pairing
symmetry is the point-contact Andreev-reflection spectroscopy (PCARS).
\cite{Hu1526, Tanaka3451,RevModPhys.77.109} The situations is still too
early to make a conclusion however. Some PCARS measurements showed two
coherent peaks and indicated that SC pairing state might be fully
gaped on the Fermi surface (FS).\cite{ PhysRevB.79.012503,
PhysRevLett.105.237002,springerlink:10.1007/s10948-009-0469-6}
Others showed a zero-bias conductance peak (ZBCP) and implied the presence of zero-energy
bound states or Andreev bound state (ABS) on the interface. \cite{0295-5075-83-5-57004,
0953-2048-21-9-092003, 0953-2048-23-5-054009} Moreover, depending on
the direction of the sample interface, some PCARSs have shown ZBCP
coexisting with finite-energy coherent peaks.
\cite{0953-2048-22-1-015018, 1367-2630-11-2-025015} Most of
theoretical studies so far have focused on the explanation of the ZBCP.
\cite{PhysRevB.79.174529, PhysRevB.79.174526}
It is important to carefully identify whether a gap node exists through the PCARS data.

In the current paper, we use a minimal two-orbital model
\cite{Raghu2008} to study the spin dynamics, charge dynamics, and
PCARS of FeSCs. Motivated by the recent STM
experiment\cite{Song2011} mentioned previously, we consider a
pairing gap of a primary $s_\pm$-wave component plus a secondary
extended $s$-wave component, called the extended $s_\pm$-wave state.
In fact, this mixed pairing state is
supported by a theoretical work of Yang {\em et al.} \cite{Yang2011}
who did a variational quantum Monte Carlo calculation and concluded
that both $s_\pm$-wave and extended $s$-wave pairings are equally
energetically favorable in the five-band FeSCs. In the studies of
spin and charge dynamics, it will be shown that in addition to
wave vectors $(\pm \pi, 0)$ that most previous works focused on, strong coherence
peaks can also occur around wave vectors $(\pm\pi/2,0)$ and $(\pm\pi,\pm\pi/2)$,
which is a unique feature in the extended $s_\pm$-wave state.
In the studies of PCARS, considering an external anisotropy effect due
to an applied bias voltage that leads to a relatively larger (smaller)
Fermi pocket for $\beta_1$- ($\beta_2$-) band and hence a node can develop in the
$(\pi,0)$ direction in $\beta_1$-band,\cite{Hung2012} low-energy
differential conductance $dI/dV$ along the [100] direction will feature
a $V$-shape curve and manifests the existence of a node.

This paper is organized as follows. In Sec.~\ref{MODEL}, we
introduce the two-orbital model and especially show how the mixed gap
behaves in different FSs as the secondary extended $s$-wave component
changes. Sec.~\ref{SPIN AND CHARGE DYNAMICS} is devoted to study the
dynamical spin and charge susceptibilities for the model introduced
in Sec.~\ref{MODEL}. Section~\ref{Point-contact
Andreev-reflection spectroscopy} gives a theoretical study of the
PCARS. Sec.~\ref{summary} is a brief summary. For self sustainability,
Appendix~\ref{A} gives detailed forms of the irreducible spin and charge response
functions of a two-band superconductor. A
brief discussion of Random-Phase Approximation (RPA) on the vertex-corrected
spin and charge response functions is given in Appendix~\ref{B}.

\section{MODEL}\label{MODEL}

We consider a minimal two-orbital model for iron-based superconductors in which
both $d_{xz}$ and $d_{yz}$ orbitals, coupled by the $d_{xy}$ orbital, are considered
in a two-dimensional square lattice. The Hamiltonian is\cite{Raghu2008}
\begin{eqnarray}
H_{0}\!=\!\sum_{\mathbf{k}\sigma}\psi^\dag_{\mathbf{k}\sigma}\left[\begin{array}{clr}
{\epsilon_{x}({\mathbf k})-\mu} & \quad{\epsilon_{xy}({\mathbf k})}\vspace{1mm}\\
{\epsilon_{xy}({\mathbf k})} & {\epsilon_{y}({\mathbf k})-\mu}
\end{array}\right]\psi_{\mathbf{k}\sigma},
\label{H0}
\end{eqnarray}
where $\psi^\dag_{\mathbf{k}\sigma}\equiv[\,d_{x\sigma}^\dag(\mathbf k),
d_{y\sigma}^\dag(\mathbf k)\,]$ with  $d_{x\sigma}^\dag(\mathbf k)$
[$d_{y\sigma}^\dag(\mathbf k)$] creating an electron in orbital
$d_{xz}$ ($d_{yz}$) of wave vector ${\bf k}$ and spin $\sigma$ and
\begin{eqnarray}
\epsilon_{x}({\mathbf k})&=&-2t_1\cos{k_x}-2t_2\cos{k_y}-4t_3\cos{k_x}\cos{k_y},\nonumber\\
\epsilon_{y}({\mathbf k})&=&-2t_2\cos{k_x}-2t_1\cos{k_y}-4t_3\cos{k_x}\cos{k_y},\nonumber\\
\epsilon_{xy}({\mathbf k})&=&-4t_4\sin{k_x}\sin{k_y}.
\label{epsilon}
\end{eqnarray}
Here $t_1$, $t_2$ are the nearest-neighbor hoppings and $t_3$, $t_4$
are the next-nearest-neighbor hoppings. After Bogoliubov
transformation, Hamiltonian (\ref{H0}) becomes
\begin{eqnarray}
H_{0}\!=\!\sum_{\mathbf{k},\sigma,\nu=\pm}\xi_{\nu}(\mathbf
k)\gamma_{\nu\sigma}^\dag(\mathbf k)\gamma_{\nu\sigma}(\mathbf k),
\label{H01}
\end{eqnarray}
where the band dispersions
\begin{eqnarray}
\xi_{\pm}(\mathbf k)=\epsilon_{+}(\mathbf k)\pm
\sqrt{\epsilon^2_{-}(\mathbf k)+\epsilon^2_{xy}(\mathbf k)}-\mu
\label{xi+-}
\end{eqnarray}
with $\epsilon_{\pm}(\mathbf k)\equiv[\epsilon_x(\mathbf k)\pm
\epsilon_y(\mathbf k)]/2$ and $\gamma_{\nu\sigma}(\mathbf k)$ is the
band operator which destroys a quasiparticle of wave vector
$\mathbf k$ and spin $\sigma$ in band $\nu$. Operators $d_{r\sigma}(\mathbf k)$
($r=x,y$) and $\gamma_{\nu\sigma}(\mathbf k)$ ($\nu=\pm$) are related to each other via the
transformation
\begin{eqnarray}
d_{r\sigma}(\mathbf k)=\sum_{\nu=\pm}a_\nu^r(\mathbf
k)\gamma_{\nu\sigma}(\mathbf k),
\label{dr}
\end{eqnarray}
where
\begin{eqnarray}
&&{a_{+}^{x}\!({\mathbf k})}\!=\!{a_{-}^{y}\!({\mathbf k})}\!=\!\text
{sgn}[\epsilon_{xy}({\mathbf
k})]\left[\frac{1}{2}+\frac{\epsilon_{-}({\mathbf
k})}{2\sqrt{\epsilon_{-}^{2}({\mathbf k})+\epsilon_{xy}^{2}({\mathbf
k})}}\right]^{1\over 2}\nonumber\\
&&{a_{+}^{y}\!({\mathbf k})}\!=\!{-a_{-}^{x}({\mathbf
k})}\!=\left[\frac{1}{2}-\frac{\epsilon_{-}({\mathbf
k})}{2\sqrt{\epsilon_{-}^{2}({\mathbf k})+\epsilon_{xy}^{2}({\mathbf
k})}}\right]^{1\over 2}.
\label{a+-}
\end{eqnarray}

\begin{figure}[t]
\vspace{0.0cm}
\includegraphics[width=0.5\textwidth]{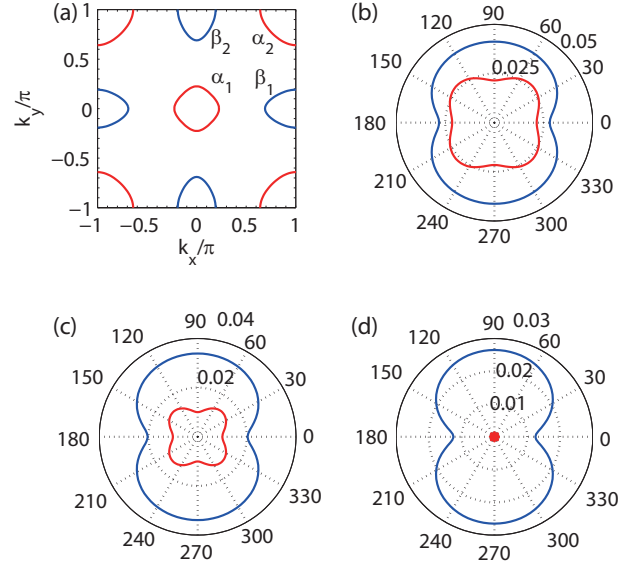}
\vspace{-1cm} \caption {(Color online) Frame (a): Fermi surfaces of the
two-band model in the unfolded BZ. Red (blue) curves correspond to
the hole (electron) Fermi pockets $\alpha_1$ and $\alpha_2$
($\beta_1$ and $\beta_2$). Frame (b)-(d): Gap magnitudes around the $\alpha_2$ (red) and
$\beta_1$ (blue) FSs for mixing ratio
(b) $x=0$, (c) $x=0.2$, and (d) $x=0.4$ respectively.}
\vspace{-0.3cm}
\label{fig1}
\end{figure}

Considering a spin-singlet pairing state, the following BCS mean-filed interaction Hamiltonian
\begin{eqnarray}
H_{\text{SC}}\!=\!-\sum_{{\mathbf k},\nu=\pm}[\Delta_\nu(\mathbf
k)\gamma_{\nu\uparrow}^\dag(\mathbf
k)\gamma_{\nu\downarrow}^\dag(-\mathbf k)+ \text{H.c.}]
\label{Hsc}
\end{eqnarray}
can be added to $H_0$ in (\ref{H01}).
We consider the gap function being in the extended $s_\pm$ state:
\begin{eqnarray}
\Delta_\nu(\mathbf{k}) = \Delta_1\cos{k_x}\cos{k_y}+{\Delta_2\over
2}\left(\cos{k_x}+\cos{k_y}\right)\qquad\nonumber\\
\equiv\Delta_0\left[\left(1-x\right)\cos{k_x}\cos{k_y}+ {x\over
2}(\cos{k_x}+\cos{k_y})\right],\, \label{Deltak}
\end{eqnarray}
where a secondary extended $s$ component ($\propto \Delta_2$) is
added to the dominant $s_\pm$ component ($\propto\Delta_1$).\cite{Song2011,Yang2011}
A ``mixing ratio" $x$ is introduced in the second line of (\ref{Deltak})
for convenience. For simplicity, the gap functions
are taken to be the same in both bands.

Fig.~\ref{fig1}(a) plots the FSs of the two-band model in which
there are two hole Fermi pockets [$\alpha_1$ and $\alpha_2$
associated with $\xi_-(\bm k_{\rm F})=0$] and two electron Fermi
pockets [$\beta_1$ and $\beta_2$ associated with $\xi_+(\bm k_{\rm
F})=0$]. The parameters used are $t_1=-0.8$, $t_2=1.4$,
$t_3=t_4=-0.9$, and chemical potential $\mu=1.15$.\cite{Hung2012}
All energies are in units of $t(>0)$ which is material dependent.
In Figs.~\ref{fig1}(b)-\ref{fig1}(d), we plot the gap magnitudes
associated with $\alpha_2$- and $\beta_1$-band FSs by varying the
mixing ratio $x$ and with a fixed amplitude $\Delta_0=0.05$
(also in units of $t$). The gaps associated with $\alpha_2$- and
$\beta_1$-band FSs are of particular interest in the current context,
which will become clear shortly. As shown in Figs.~\ref{fig1}(b)-\ref{fig1}(d),
gap magnitude on the $\alpha_2$ FS decreases as the mixing ratio $x$ increases.
In particular, in the case of $x=0.4$ [see Fig.~\ref{fig1}(d)], the gap
magnitude around $\alpha_2$ band can reduce to zero ({\em i.e.},
becomes nodal). Note that the gap magnitude around the $\alpha_1$ band
(not shown) remains roughly unchanged as $x$ changes. In addition,
of equal importance, the gap magnitude around the $\beta_1$-band FS becomes smaller and
smaller in the $k_x$ axis as $x$ increases. When $x$ is increased to be roughly 0.7,
or when there is an induced band anisotropy due to an applied voltage
with $x=0.4$ (see Fig.~\ref{fig5}), a gap node
can actually develop in the $\beta_1$ Fermi pocket in the $k_x$ direction.

\section{SPIN AND CHARGE DYNAMICS}\label{SPIN AND CHARGE DYNAMICS}

This section is devoted to study the dynamical spin and charge susceptibilities
of FeSCs. To save the space here and still for self sustainability, detailed forms of
the irreducible spin and charge response functions of a two-band superconductor
are to be given in Appendix~\ref{A}. A brief
discussion of the Random-Phase Approximation (RPA) on the vertex-corrected spin
and charge response functions is given in Appendix~\ref{B}.

\begin{figure}[t]
\vspace{0.0cm}
\includegraphics[width=0.5\textwidth]{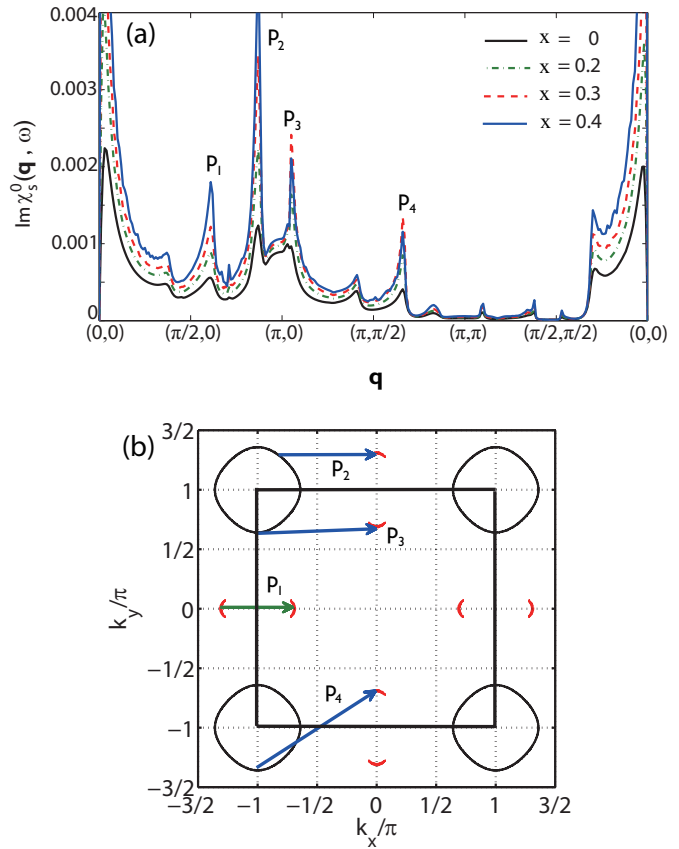}
\vspace{-0.4cm} \caption {(Color online) (a) Irreducible dynamic spin
susceptibility ${\rm Im}\chi_{s}^{0}(\mathbf
{q},\omega=0.3\Delta_0)$ for wavevectors ${\bf q}$ along the
$(0,0)\rightarrow(\pi,0)\rightarrow(\pi,\pi)\rightarrow(0,0)$ route
in the BZ and for different mixing ratio $x$.
(b) Contour plot of the excitation spectrum $E_{\bf k}=\omega/2=0.15\Delta_0$ with
$x=0.4$. Blue arrows ($p_2$--$p_4$) correspond to nesting wave vectors between the hole
$\alpha_2$ (black) and electron $\beta_1$ (red) bands. The one green arrow ($p_1$)
corresponds to a nesting wave vector within the electron $\beta_1$ band.
These vectors correspond to those peaks emerging in frame (a).}
\vspace{-0.2cm} \label{fig2}
\end{figure}

\subsection{Dynamic Spin Susceptibility}
\label{Dynamic Spin Susceptibility}

We first consider the dynamical spin susceptibility in the $T\rightarrow 0$ limit.
For spin-singlet pairing, dynamical spin susceptibility is proportional to the
imaginary part of the spin response function defined as
\begin{eqnarray}
\chi_{+-}(\mathbf q,i\omega_n)=\sum_{r,t}\int_0^\beta \text{d}\tau
\text{e}^{i\omega_n\tau}\langle{T_\tau S_r^+ (\mathbf
q,\tau)S_t^-(-\mathbf q,0)}\rangle,\nonumber\\
\label{chi+-}
\end{eqnarray}
where $r,t=x,y$ are the orbital indices and the spin operator
\begin{eqnarray}
S_r^+(\mathbf q,\tau)&=&\sum_{\mathbf k}d_{r\uparrow}^\dag(\mathbf
k+\mathbf q,\tau)d_{r\downarrow}(\mathbf k,\tau)\nonumber\\
&=&[S_r^-(-\mathbf q,\tau)]^\dagger.
\label{S+-}
\end{eqnarray}
For simplicity, we denote $\chi_s\equiv\chi_{+-}$ throughout this paper.
In the $T\rightarrow 0$ limit,
only the first and second terms contribute in the irreducible spin response
function $\chi_s^0$ given in Eq.~(\ref{Eq:chi_s^0}) in Appendix~\ref{A}.
Furthermore, only the first term contributes if $\omega>0$ is concerned.
Consequently, apart from the effect of the coherence factors, the contributions to
${\rm Im}\chi_s^0$ are dominated by the condition of the particle-particle excitation
\begin{eqnarray}
\omega=E_{\nu'}(\mathbf {k+q})+E_\nu(\mathbf {k}).
\label{Eq:resonance}
\end{eqnarray}
Here $E_{\nu}({\mathbf k})=\sqrt{\xi_{\nu}^2({\mathbf
k})+|\Delta_\nu({\mathbf k})|^{2}}$ is the quasiparticle excitation
spectrum with the band indix $\nu=+,-$.

In Fig.~\ref{fig2}(a), we show the calculating results of the irreducible spin
susceptibility ${\rm Im}\chi_s^0$ for four different mixing ratio $x$
and a fixed (low) energy $\omega=0.3\Delta_0$.
This two-dimensional plot is made by connecting important symmetry points
in the BZ. As shown in Fig.~\ref{fig2}(a), ${\rm Im}\chi_s^0$ has
strong peaks emerging at wave vectors near $(\pi/2,0)$, $(\pi,0)$, and
$(\pi,\pi/2)$. Of most interest, with the same broadening $\delta=0.01$
(in units of $t$) and the same gap amplitude $\Delta_0$ used in all cases,
the peaks are seen to be much stronger in the
larger $x$ cases compared to those in the pure $s_\pm$-wave
($x=0$) case. To understand these phenomena, one just considers the low-energy
particle-particle excitations with respect to the gap mixing. For pure $s_\pm$-wave
pairing ($x=0$) case, Fermi surfaces are fully gapped and consequently
low-energy particle-particle excitation is suppressed. In contrast, for extended
$s_\pm$-wave pairing of larger $x$, gap nodes can develop around the
$\alpha_2$-band FSs as well as in certain ${\bf k}$ areas around the $\beta$-band
FSs. Consequently low-energy particle-particle excitation
is enabled. (Note that no low-energy excitation with $\omega<\Delta_0$ is allowed
in connection with the fully gapped $\alpha_1$-band FS regardless of what the
mixing ratio $x$ is.)

To understand the origins of the peaks further, in Fig.~\ref{fig2}(b) we plot energy contours of the
quasiparticle excitation $E_{\bf k}=\omega/2=0.15\Delta_0$ with $x=0.4$.
The arrows denote the nesting wave vectors for which degenerate strong two-particle
excitations are associated with. These particular wave vectors, labeled by $p_1\sim p_4$,
are exactly those that the strong peaks are associated with in Fig.~\ref{fig2}(a).
As a matter of the fact, strong peaks emerge due to the scattering
between electron and hole Fermi pockets (blue vectors) as well as
within the same electron Fermi pocket (green vector).

\begin{figure}[t]
\vspace{0.0cm}
\includegraphics[width=0.5\textwidth]{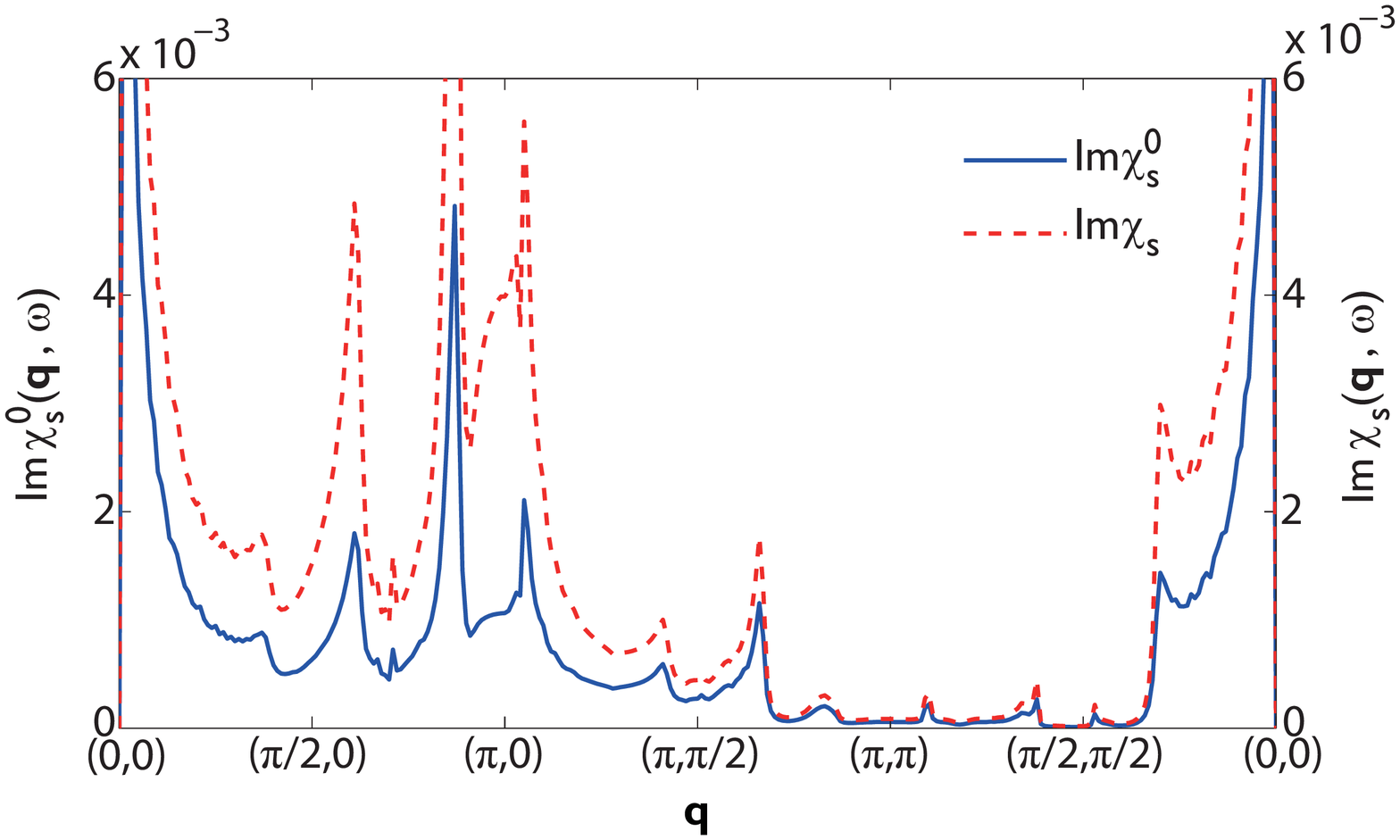}
\vspace{-0.6cm} \caption {(Color online) Comparison of the spectra in
irreducible and RPA-corrected spin susceptibility ${\rm Im}\chi_{s}^0$ and ${\rm
Im}\chi_s$ for $x=0.4$ and $\omega=0.3\Delta_0$. Coulomb repulsion
is taken to be $U=1.8$ for ${\rm Im}\chi_s$.} \label{fig3}
\end{figure}

To see how the behaviors of the full spin susceptibility differ from those
of the irreducible one, in Fig.~\ref{fig3} we compare the results of
${\rm Im}\chi_s^0$ and ${\rm Im}\chi_s$ for $x=0.4$ and $\omega=0.3\Delta_0$.
Here the full spin response function $\chi_s$ is calculated under RPA. A brief discussion
on how $\chi_s$ is obtained is given in Appendix~\ref{B}.
In calculating $\chi_s$, we have set the intro-orbital Coulomb repulsion $U=1.8$
(in units of $t$) and the Hund's coupling $J=0$ (see Appendix~\ref{B}).
In view of Fig.~\ref{fig3}, while enhancement
occurs for ${\rm Im}\chi_s$, the results are qualitatively similar
between ${\rm Im}\chi_s^0$ and ${\rm Im}\chi_s$.
Concerning the results of ${\rm Im}\chi_s^0$ and/or ${\rm Im}\chi_s$ in
Fig.~\ref{fig3}, we note the following two facts. First, the spectra for
a higher (lower) energy with $\omega >0.3\Delta_0$ ($\omega <0.3\Delta_0$)
are basically the same as those of the $\omega =0.3\Delta_0$ case.
The only major difference is that the intensity of the peaks is stronger (weaker)
in the higher (lower) energy case.
Secondly, in the parameter regime we are studying, the peak around ${\mathbf q}=(\pi,0)$
is enhanced most in ${\rm Im}\chi_s$. Perhaps this
is the case of most interest as most current
INS experiments have focused on the possible magnetic peaks around ${\mathbf q}=(\pi,0)$.
\cite{Shamoto2010,Christianson2008,Ishikado2011,Taylor2011,Qiu2009,Mook2010}
We emphasize that strong magnetic peaks can also emerge at wave vectors around
$(\pi/2,0)$ and $(\pi,\pi/2)$ if an extended $s_\pm$-wave pairing state of a
larger mixing ratio $x$ is present.

\begin{figure}[t]
\vspace{0.0cm}
\includegraphics[width=0.5\textwidth]{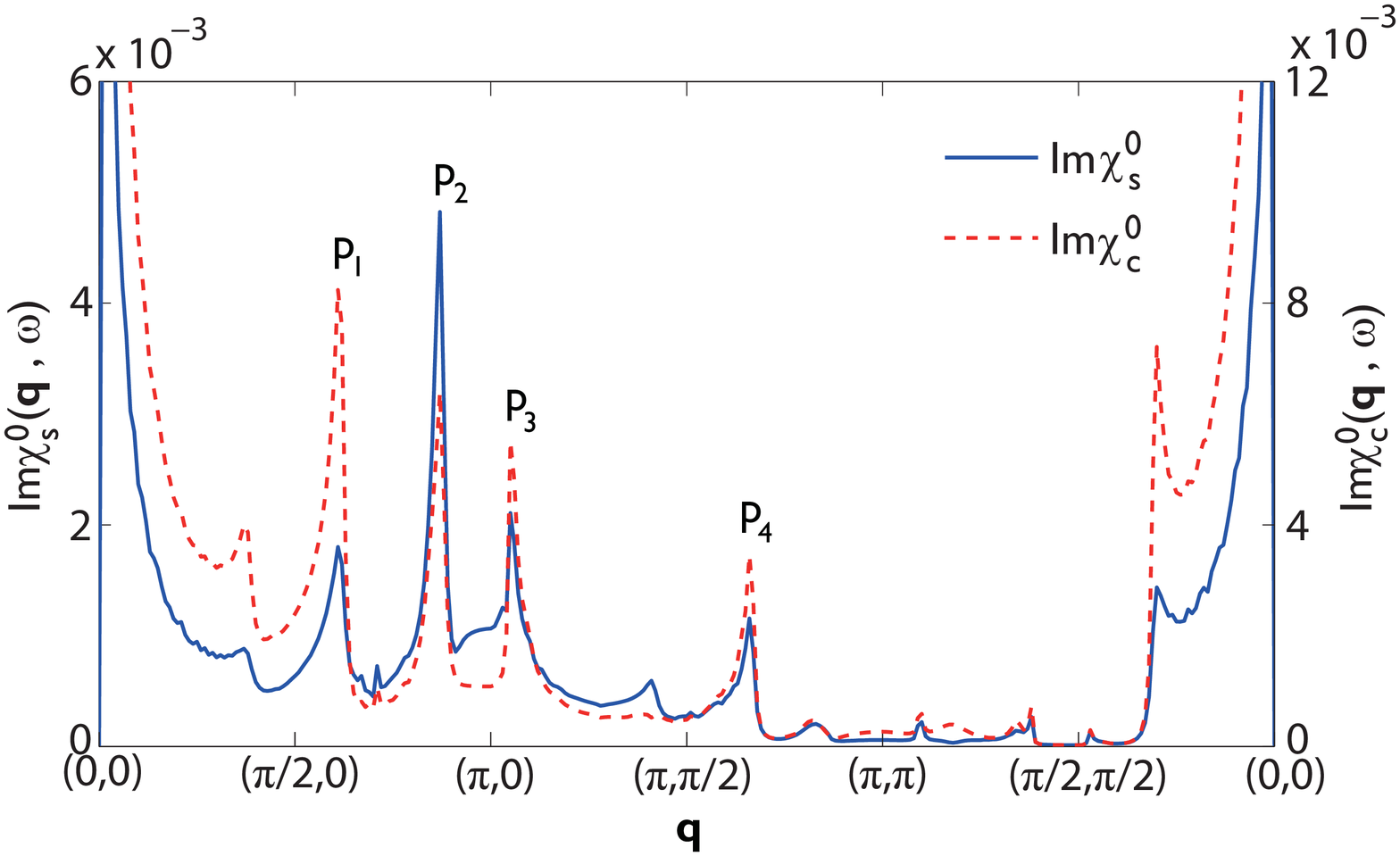}
\vspace{-0.6cm} \caption {(Color online) Comparison of the spectra in
irreducible spin and charge susceptibilities ${\rm Im}\chi_{s}^0$ and
${\rm Im}\chi_c^0$ for $x=0.4$ as $\omega=0.3\Delta_0$.}
\label{fig4}
\end{figure}

\subsection{Dynamic Charge Susceptibility}
\label{Dynamic Charge Susceptibility}

We now turn to study the dynamical charge susceptibility
in the $T\rightarrow 0$ limit. Dynamical charge susceptibility is proportional to the
imaginary part of the charge response function
\begin{eqnarray}
\chi_{c}(\mathbf q,i\omega_n)=\sum_{r,t}\int_0^\beta \text{d}\tau
\text{e}^{i\omega_n\tau}\langle{T_\tau \rho_r(\mathbf
q,\tau)\rho_t(-\mathbf q,0)}\rangle,\quad
\label{chic0}
\end{eqnarray}
where the density operator
\begin{eqnarray}
\rho_r(\mathbf q,\tau)=\sum_{\mathbf
k\sigma}d_{r\sigma}^\dag(\mathbf k+\mathbf
q,\tau)d_{r\sigma}(\mathbf k,\tau).
\label{rhor}
\end{eqnarray}
The detailed form of the irreducible charge response function, $\chi_{c}^0$, is given in
Eq.~(\ref{Eq:chi_c^0}) in Appendix~\ref{A}. In Fig.~\ref{fig4}, we
show and compare the spectra of the irreducible charge and spin
susceptibilities ${\rm Im}\chi_c^0$ and ${\rm Im}\chi_s^0$ for $x=0.4$
and $\omega=0.3\Delta_0$. One sees clearly that
similar strong peaks emerge at wave vectors near $(\pi/2,0)$, $(\pi,\pi/2)$
as well as ($\pi,0$) in both ${\rm Im}\chi_s^0$ and ${\rm Im}\chi_c^0$.
Therefore the charge peaks with wave vectors near
$(\pi/2,0)$, $(\pi,\pi/2)$, and ($\pi,0$) can also be considered as the
signature for the extended $s_\pm$-wave pairing state.
One is noted a key difference in the spectra of ${\rm Im}\chi_s^0$ and
${\rm Im}\chi_c^0$ however. In ${\rm Im}\chi_s^0$, the strongest
peak is associated with wavevector $p_2$, while in ${\rm Im}\chi_c^0$,
the strongest peak is associated with wavevector $p_1$.
The difference is due to different coherence factors
coupled to $\chi_c^0$ and $\chi_s^0$ [see the first term of
$\chi_s^0$ and $\chi_c^0$ in Eqs.~(\ref{Eq:chi_s^0}) and
(\ref{Eq:chi_c^0})].

\section{Point-contact Andreev-reflection spectroscopy}
\label{Point-contact Andreev-reflection spectroscopy}

When a free electron is injected from a normal metal into a
superconductor (forming a interface at $x=0$) with an incident angle
$\theta$, there are two possible reflections. One is the  normal
reflection (reflected as electrons) and
another is the Andreev reflection (reflected as holes, due to
electron and hole coupling in the $\mathbf{k}$ subspace).
The resulting wave function $\Psi_N$ in the normal side ($x<0$)
can be obtained by the
superposition of the above two kinds of reflected waves and the
incident wave. In the superconducting side ($x>0$), the resulting
wave function $\Psi_S$ is the superposition of the electron-like and
hole-like quasiparticle wave functions, which can be obtained by
solving the Andreev equation.

On key issue for the superconductor is that electron-like and
hole-like quasiparticles will experience the effective
pairing potential $\Delta(\theta)$ and $\Delta(\pi-\theta)$ respectively.
As a matter of the fact, resulting tunneling conductance will depend
strongly on the pairing symmetry and the direction of the tunneling current.
Considering tunneling current along
the [100] direction, $\Delta(\theta)=\Delta(\pi-\theta)$ for the current extended
$s_\pm$-wave pairing and tunneling conductance will feature a nodeless gap if
the mixing ratio $x$ is small. However, as mentioned before, when
$x$ is large or when there is a large band anisotropy occurring
due to an applied voltage (see later), the gap associated with the $\beta_1$-band
will shrink in the $k_x$ direction and could eventually develop a node in such direction.
This nodal gap will then be manifested in the tunneling conductance along such direction.

By matching the normal and superconducting wave functions and their derivatives at the
interface $x=0$:
\begin{eqnarray}
\psi _{N}\left( \mathbf{r}\right) |_{x=0^{-}} &&=\psi _{S}\left( \mathbf{r}%
\right) |_{x=0^{+}},  \label{eq:boundary conditions} \\
\frac{2mH}{\hbar ^{2}}\psi _{S}\left( \mathbf{r}\right) |_{x=0^{+}}&&=\frac{%
d\psi _{S}\left( \mathbf{r}\right) }{dx}|_{x=0^{+}}-\frac{d\psi _{N}\left(%
\mathbf{r}\right) }{dx}|_{x=0^{-}},\nonumber
\end{eqnarray}%
where $H$ denotes the strength of a delta-function like barrier potential
$V(x)\equiv H\delta(x)$, one can solve the normal and Andreev
reflection coefficients $r_N$ and $r_A$.
With the solved $r_N$ and $r_A$ and based on the well-known formula
for the conductance $\sigma_S=1+|r_A|^2-|r_N|^2$
(see, for example, Ref.~[\onlinecite{PhysRevB.25.4515}]),
the normalized differential conductance $dI/dV$ can be obtained to be
\begin{equation}
\tilde{\sigma}_{S}\left( E, \theta\right)= \frac{16\left( 1+ \Gamma
^{2}\right)\cos^4 \theta  +4Z^{2}\left( 1-\Gamma ^{2}\right) \cos^2
\theta }{\left\vert 4 \cos^2 \theta+Z^{2}(1- \Gamma^2 ) \right\vert^{2}},
\label{the tunneling conductance}
\end{equation}
where $Z=2mH/\hbar^2$ is the effective potential barrier and
\begin{eqnarray}
\Gamma\equiv\frac{E}{\left\vert \Delta(\theta)\right\vert }-\sqrt{\left(
\frac{E}{\left\vert \Delta(\theta)\right\vert }\right) ^{2}-1}.
\label{eq:Gamma}
\end{eqnarray}
In a real experimental setup, the above derivation is equivalent to consider
an STM tip made of a normal metal and being well point-contacted to the superconductor.
Similar derivations of Eqs.~(\ref{the tunneling conductance}) and
(\ref{eq:Gamma}) can be found in Ref.~[\onlinecite{Tanaka3451}] where
a single-band system was considered.

\begin{figure}[t]
\vspace{0.0cm}
\includegraphics[width=8cm]{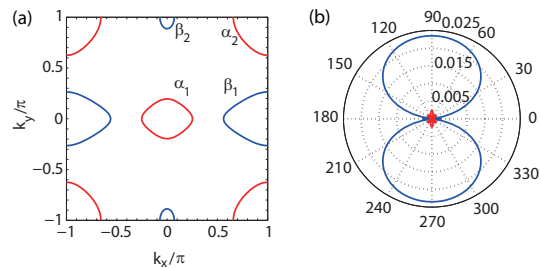}
\vspace{-0.2cm} \caption {(Color online) (a) Fermi surfaces of the band
dispersions given in (\ref{BandWithAnisotropy}). The parameters are
the same as those in Fig.~\ref{fig1}(a) together with an anisotropy $\delta\epsilon=0.3$.
(b) Gap magnitudes of the extended $s_\pm$-wave gap (\ref{Deltak}) associated with
$\alpha_2$- (red) and $\beta_1$-band (blue) FSs with $x=0.4$ and $\Delta_0=0.05$.}
\label{fig5}
\end{figure}

In the current context, for simplicity, we assume that the two
bands are decoupled completely. Nevertheless the resulting tunneling
conductance is considered to be contributed by all $\alpha_1$, $\alpha_2$, $\beta_1$, and
$\beta_2$ FSs. To see the possible nodal effect for the conductance current
along the [100] direction, we consider
a band anisotropy effect induced by an applied voltage.
The applied voltage should be much larger than the pairing gap magnitude.\cite{Hung2012}
Consequently the induced anisotropic band dispersion can be approximated by
\begin{eqnarray}
\tilde{\xi}_{\pm}(\mathbf k)=\epsilon_{+}(\mathbf k)\pm
\sqrt{(\epsilon_{-}(\mathbf
k)+\delta\epsilon)^2+\epsilon^2_{xy}(\mathbf k)}-\mu,
\label{BandWithAnisotropy}
\end{eqnarray}
where $\delta\epsilon$ denotes the induced anisotropy.
In Fig.~\ref{fig5}(a), FSs of the anisotropic band (\ref{BandWithAnisotropy})
are shown in which $\beta_1$-band Fermi pocket becomes larger, while
$\beta_2$-band Fermi pocket becomes smaller as compared to those in
Fig.~\ref{fig1}(a). The parameters used in Fig.~\ref{fig5}(a) are
the same as those in Fig.~\ref{fig1}(a), {\em i.e.}, $t_1=-0.8$, $t_2=1.4$,
$t_3=t_4=-0.9$, and $\mu=1.15$. In addition,the induced anisotropy is chosen to be
$\delta\epsilon=0.3$. Fig.~\ref{fig5}(b) shows how the extended $s_\pm$-wave gap
behaves in the $\alpha_2$- and $\beta_1$-band FSs with $x=0.4$ and
$\Delta_0=0.05$. The most important feature in these gaps is that
gap nodes develop in both $\alpha_2$- and $\beta_1$-band FSs.
For the $\alpha_2$-band gap, the node appears to be nodal almost
everywhere around the $\alpha_2$ FS. In contrast,
for the $\beta_1$-band gap, the node appears in the $k_x$ direction only which
should have a strong effect on the conductance current along the [100] direction.

\begin{figure}[t]
\vspace{0.0cm}
\includegraphics[width=8cm]{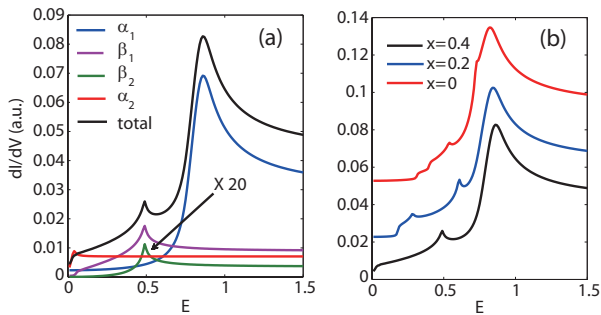}
\vspace{-0.2cm} \caption {(Color online) (a) Resulting [100] differential
conductance $dI/dV$ vs. tunneling voltage $E$ (black line).
Parameters are the same as those in Fig.~\ref{fig5}.
For comparison, contributions due to four Fermi pockets are plotted respectively
in which $\beta_2$ contribution is magnified by a factor of 20 for a better view.
(b) Comparison of resulting [100] $dI/dV(E)$ curves with $x=0.4$, $0.2$, and $0$.
For clarity the curve of $x=0.2$ (0) is shifted by 0.02 (0.04) vertically.}
\label{fig6}
\end{figure}

Figure~\ref{fig6}(a) plots the [100] differential conductance $dI/dV$
versus the tunneling voltage $E$ for a normal-superconducting junction. The superconductor
is made of the two-band iron-based with the induced anisotropic
band dispersions given in (\ref{BandWithAnisotropy}) and the extended
$s_\pm$-wave pairing gap given in (\ref{Deltak}).
The parameters used are the same as those in Fig.~\ref{fig5}. In addition,
the strength of barrier is taken to be $H=10$ (in units of $t$).
To be more explicitly, $dI/dV(E)$ is obtained by
integrating $\tilde{\sigma}_{S}(E,\theta)$ in
(\ref{the tunneling conductance}) over the $\theta$ angle from $-\pi/2$ to
$\pi/2$. For comparison purpose, we have plotted the contribution respectively
due to $\alpha_1$, $\alpha_2$, $\beta_1$, and $\beta_2$ bands.
Of most interest is that, in the low $E$ regime, in contrast to $\alpha_1$,
$\alpha_2$, and $\beta_2$ bands whose contribution is either little or almost
independent of $E$, the contribution due to $\beta_1$ band is dominant and
linear in $E$. This can be understood in the following.
$\beta_2$-band is located far away from the $k_x$ axis,
thus its contribution (green line) to the differential conductance is the least.
For the $\alpha_2$-band, the gap is nodal almost everywhere on the FS,
so its contribution to $dI/dV(E)$ (red line) is almost independent of $E$ ({\em i.e.}, obeying the
Ohm's law). For the $\alpha_1$-band, finite $s$-wave gap is fully developed
in its FS, consequently its corresponding $dI/dV(E)$ (blue line) exhibits a $U$ shape.
In contrast for the $\beta_1$-band, gap node develops in the $k_x$ direction,
so for the [100] differential conductance, it exhibits a $V$ shape (purple line)
and manifests the existence of a gap node along such direction.
The overall $V$-shape $dI/dV(E)$ at low $E$ seems to be in consistence
with the STM measurement reported in Ref.~[\onlinecite{Song2011}].

In Fig.~\ref{fig6}(b), we show the [100] $dI/dV$ versus $E$ by changing the gap
mixing ratio $x$. It is evident that when $x=0$ for a pure $s_\pm$-wave pairing,
all FSs are fully gapped and consequently the resulting $dI/dV(E)$
exhibits a $U$-shape behavior.
Those tiny peaks corresponds to quasiparticle excitations associated
with different Fermi surfaces.
It is in great contrast to the case of $x=0.4$ in which $dI/dV(E)$ exhibits a
$V$-shape behavior, as already shown in Fig.~\ref{fig6}(a).

\section{summary}\label{summary}

In this paper, motivated by the recent experiment of Song \emph{et al.}
\cite{Song2011}, we intend to study the spin dynamics, charge
dynamics, and point-contact Andreev-reflection spectroscopy (PCARS)
of iron-based superconductors of a possible extended $s_\pm$-wave pairing
symmetry. We consider a minimal two-band model in which
superconducting gap is dominated by the $s_\pm$ component but mixed
by a secondary extended $s$ component. Comparing to the pure
$s_\pm$-wave gap of finite gaps around all four Fermi pockets, in the
present extended $s_\pm$-wave pairing state, gap nodes can develop in the
Fermi pockets near the zone corner and/or the zone boundary.
Useful signatures associated with the possible nodes are identified in
spin dynamics, charge dynamics, as well as in the PCARS.
It is hoped that the features discussed in this paper can be carefully
investigated by experiments to help sorting out the pairing symmetry of
iron-based superconductors.

Finally we remark that some recent inelastic neutron scattering
experiments on $\text{K}_x\text{Fe}_{2}\text{Se}_2$ samples have
uncovered magnetic resonance peak at $\textbf{q}=(\pi,\pi/2)$ in
unfolded BZ.\cite{Friemel2012} This implies that the pairing state
in these materials is likely to be $d$-wave. Fermi surface topology
in these materials is, in fact, quite different from the iron-based
superconductors that we are considering.

\acknowledgements

This work was supported by National Science Council of Taiwan (Grant No.
NSC 99-2112-M-003 -006 -MY3). CSL was supported by Hebei Provincial Natural
Science Foundation of China (Grant No. A2010001116) and the National
Natural Science Foundation of China (Grant No. 10974169).
We also acknowledge the support from the National
Center for Theoretical Sciences, Taiwan.

\appendix
\section{Irreducible Spin and Charge Response Functions}\label{A}

Within the two-orbital model proposed by Raghu {\it et
al.},\cite{Raghu2008} irreducible spin and charge response functions can be
derived to be (see also the derivation in Ref.~[\onlinecite{Masuda2010}])
\begin{widetext}
\begin{eqnarray}
{\chi_{s}^0}({\mathbf q},\omega)&=&-{\frac{1}{2N}}\sum_{\mathbf
k}\sum_{r,t}\sum_{\nu,\nu'=\pm}M^{\nu\nu'}_{rt}({\mathbf k,\mathbf
q})\left[ {A^{-}_{\mathbf k,\mathbf
q}{\frac{1-f(E_{\nu'}({\mathbf k}))-f(E_{\nu}({\mathbf k+\mathbf
q}))}{\omega-E_{\nu'}({\mathbf k})-E_{\nu}({\mathbf k+\mathbf
q})+i\delta}}} \right.
-B^{-}_{\mathbf k,\mathbf q}{\frac{1-f(E_{\nu'}({\mathbf k}))-f(E_{\nu}({\mathbf k+\mathbf q}))}{\omega+E_{\nu'}({\mathbf k})+E_{\nu}({\mathbf k+\mathbf q})+i\delta}}\nonumber\\
&-&C^{+}_{\mathbf k,\mathbf q}{\frac{f(E_{\nu'}({\mathbf
k}))-f(E_{\nu}({\mathbf k+\mathbf q}))}{\omega-E_{\nu'}({\mathbf
k})+E_{\nu}({\mathbf k+\mathbf q})+i\delta}} +\left. {D^{+}_{\mathbf
k,\mathbf q}{\frac{f(E_{\nu'}({\mathbf k}))-f(E_{\nu}({\mathbf
k+\mathbf q}))}{\omega+E_{\nu'}({\mathbf k})-E_{\nu}({\mathbf
k+\mathbf q})+i\delta}}}
\right]\equiv\sum_{rt}\chi_{s,rt}^0({\mathbf q},\omega)
\label{Eq:chi_s^0}
\end{eqnarray}
\begin{eqnarray}
{\chi_c^0}({\mathbf q},\omega)&=&-{\frac{1}{N}}\sum_{\mathbf
k}\sum_{r,t}\sum_{\nu,\nu'=\pm}M^{\nu\nu'}_{rt}({\mathbf k,\mathbf q})\left[ {A^{+}_{\mathbf k,\mathbf q}{\frac{1-f(E_{\nu'}({\mathbf k}))-f(E_{\nu}({\mathbf k+\mathbf q}))}{\omega-E_{\nu'}({\mathbf k})-E_{\nu}({\mathbf k+\mathbf q})+i\delta}}} \right.-B^{+}_{\mathbf k,\mathbf q}{\frac{1-f(E_{\nu'}({\mathbf k}))-f(E_{\nu}({\mathbf k+\mathbf q}))}{\omega+E_{\nu'}({\mathbf k})+E_{\nu}({\mathbf k+\mathbf q})+i\delta}}\nonumber\\
&-&C^{-}_{\mathbf k,\mathbf q}{\frac{f(E_{\nu'}({\mathbf
k}))-f(E_{\nu}({\mathbf k+\mathbf q}))}{\omega-E_{\nu'}({\mathbf
k})+E_{\nu}({\mathbf k+\mathbf q})+i\delta}}+\left. {D^{-}_{\mathbf
k,\mathbf q}{\frac{f(E_{\nu'}({\mathbf k}))-f(E_{\nu}({\mathbf
k+\mathbf q}))}{\omega+E_{\nu'}({\mathbf k})-E_{\nu}({\mathbf
k+\mathbf q})+i\delta}}}
\right]\equiv\sum_{rt}\chi_{c,rt}^0({\mathbf q},\omega),
\label{Eq:chi_c^0}
\end{eqnarray}
\end{widetext}
where $r,t=x,y$ correspond to the orbital indices, $\nu,\nu'=\pm$
correspond to the band indices, $N$ is the
renormalization factor, and $\delta$ is the broadening.
$f(E_{\nu}({\mathbf k}))$ is the Fermi distribution function with
$E_{\nu}({\mathbf k})=\sqrt{\xi_{\nu}^2({\mathbf
k})+|\Delta({\mathbf k})|^{2}}$ the quasiparticle excitation
spectrum, $\xi_\nu({\bf k})$ the band dispersion, and
$\Delta_\nu({\bf k})$ the superconducting gap.
The coherence factors in (\ref{Eq:chi_s^0}) and
(\ref{Eq:chi_c^0}) are respectively

\begin{eqnarray*}
{A^{\pm}_{\mathbf k,\mathbf q}}&\equiv&{\frac{1}{2}}\left[
{\left(1+\frac{\xi_{\nu'}({\mathbf k})}{E_{\nu'}({\mathbf
k})}\right)\left(1-\frac{\xi_{\nu}({\mathbf k+\mathbf
q})}{E_{\nu}({\mathbf k+\mathbf q})}\right)}
\right.\nonumber\\&\pm&\left. {\frac{\Delta_{\nu'}({\mathbf
k})\Delta_{\nu}({\mathbf k+\mathbf q})}{E_{\nu'}({\mathbf
k})E_{\nu}({\mathbf k+\mathbf q})}} \right]
\end{eqnarray*}
\begin{eqnarray}
{B^{\pm}_{\mathbf k,\mathbf q}}&\equiv&{\frac{1}{2}}\left[
{\left(1-\frac{\xi_{\nu'}({\mathbf k})}{E_{\nu'}({\mathbf
k})}\right)\left(1+\frac{\xi_{\nu}({\mathbf k+\mathbf
q})}{E_{\nu}({\mathbf k+\mathbf q})}\right)}
\right.\nonumber\\&\pm&\left. {\frac{\Delta_{\nu'}({\mathbf
k})\Delta_{\nu}({\mathbf k+\mathbf q})}{E_{\nu'}({\mathbf
k})E_{\nu}({\mathbf k+\mathbf q})}} \right]\nonumber\\
{C^{\pm}_{\mathbf k,\mathbf q}}&\equiv&{\frac{1}{2}}\left[
{\left(1+\frac{\xi_{\nu'}({\mathbf k})}{E_{\nu'}({\mathbf
k})}\right)\left(1+\frac{\xi_{\nu}({\mathbf k+\mathbf
q})}{E_{\nu}({\mathbf k+\mathbf q})}\right)}
\right.\nonumber\\&\pm&\left. {\frac{\Delta_{\nu'}({\mathbf
k})\Delta_{\nu}({\mathbf k+\mathbf q})}{E_{\nu'}({\mathbf
k})E_{\nu}({\mathbf k+\mathbf
q})}}\right]\nonumber\\
{D^{\pm}_{\mathbf k,\mathbf
q}}&\equiv&{\frac{1}{2}}\left[ {\left(1-\frac{\xi_{\nu'}({\mathbf
k})}{E_{\nu'}({\mathbf k})}\right)\left(1-\frac{\xi_{\nu}(\mathbf
k+\mathbf q)}{E_{\nu}(\mathbf k+\mathbf q)}\right)}
\right.\nonumber\\&\pm&\left. {\frac{\Delta_{\nu'}({\mathbf
k})\Delta_{\nu}({\mathbf k+\mathbf q})}{E_{\nu'}({\mathbf
k})E_{\nu}({\mathbf k+\mathbf q})}}\right]
\label{Coherence Factors}
\end{eqnarray}
and the function
\begin{eqnarray}
M^{\nu\nu'}_{rt}({\mathbf k,\mathbf q})\equiv{a_{\nu}^{r}({\mathbf
k+\mathbf q})a_{\nu'}^{r}({\mathbf k})a_{\nu'}^{t}({\mathbf
k})a_{\nu}^{t}({\mathbf k+\mathbf q})},\label{Eq:bandeffect}
\end{eqnarray}
with $a_{\nu}^{r}({\mathbf k})$ defined in Eq.~(\ref{a+-}).

\section{Random-Phase Approximation}\label{B}

Here we consider the full spin and charge response functions taking
into account the vertex correction due to various interactions. For
$d_{xz}$- and $d_{yz}$-orbital electrons, we consider the following
on-site interaction
\begin{eqnarray}
H_{\rm I}&=&U\sum_{ir}n_{i,r\uparrow}n_{i,r\downarrow}
+{\frac{V}{2}}\sum_{i,r,t\neq r}n_{ir}n_{it}\nonumber\\
&-&{\frac{J}{2}}\sum_{i,r,t\neq r}{\bf S}_{ir}\cdot{\bf S}_{it}\nonumber\\
&+&{\frac{J^\prime}{2}}\sum_{i,r,t\neq
r}\sum_{\sigma}d_{ir\sigma}^{\dag}d_{ir\bar{\sigma}}^{\dag}d_{it\bar{\sigma}}d_{it\sigma},
\label{Eq:Hamiltonian_int}
\end{eqnarray}
where $U$ and $V$ correspond to the intra- and inter-orbital Coulomb
repulsion respectively, $J$ is the Hund's coupling, and $J^\prime$
is responsible for pair hopping. Note in (\ref{Eq:Hamiltonian_int}), for
each site $i$, $n_{ir}=n_{i,r\uparrow}+n_{i,r\downarrow}$. Within
Random-Phase Approximation, irreducible spin response function in
(\ref{Eq:chi_s^0}) will be renormalized to be
\begin{eqnarray}
\hat{\chi}_{s}({\mathbf q},\omega)=\hat{\chi}_{s}^0({\mathbf
q},\omega)[{\bf I}-{\bf \Gamma}_s\hat{\chi}_{s}^{0}({\mathbf
q},\omega)]^{-1}, \label{Eq:RPA_chi_s}
\end{eqnarray}
where $\bf I$ is the $2\times 2$ unit matrix, $\hat{\chi}_{s}^0$ is
a $2\times 2$ matrix formed by the intra- and inter-orbital spin
susceptibilities defined in the last term in (\ref{Eq:chi_s^0}), {\em i.e.},
$[\hat{\chi}_{s}^0]_{rt}\equiv\chi_{s,rt}^0$, and the vertex matrix
\begin{eqnarray}
{\bf \Gamma}_s=\left( {\begin{array}{*{20}{c}}
   {U} & {J/2}  \\
   {J/2} & {U}  \\
\end{array}} \right).
\label{Eq:vertex_s}
\end{eqnarray}
While the irreducible charge response function in (\ref{Eq:chi_c^0}) will be renormalized to be
\begin{eqnarray}
\hat{\chi}_{c}({\mathbf q},\omega)=\hat{\chi}_{c}^0({\mathbf
q},\omega)[{\bf I}+{\bf \Gamma}_c\hat{\chi}_{c}^{0}({\mathbf
q},\omega)]^{-1},\label{Eq:RPA_chi_c}
\end{eqnarray}
where $[\hat{\chi}_c^0]_{rt}\equiv\chi_{c,rt}^0$ and
\begin{eqnarray}
{\bf \Gamma}_c=\left( {\begin{array}{*{20}{c}}
   {U} & {2V}  \\
   {2V} & {U}  \\
\end{array}} \right).
\label{Eq:vertex_c}
\end{eqnarray}
In the calculation of ${\rm Im}\chi_s$ in Sec.~III, we have chosen $U=1.8$ and $J=0$
for simplicity. We do not actually calculate the RPA-corrected ${\rm Im}\chi_c$
in the current context.


%

\end{document}